\documentclass[%
showpacs,
preprintnumbers,
amsmath,
amssymb,
pre,
twocolumn,
longbibliography
]{revtex4-2}

\usepackage{graphicx}
\usepackage{dcolumn}
\usepackage{bm}
\usepackage[dvipdfmx]{color,hyperref}
\hypersetup{
    colorlinks=true,
    citecolor=blue,
    linkcolor=blue,
    urlcolor=blue,
}
\newcommand{\bB}{\bm{B}}
\newcommand{\bF}{{\bf F}}

\newcommand{\bzeta}{{\boldsymbol{\zeta}}}
\newcommand{\bmeta}{{\boldsymbol{\eta}}}
\newcommand{\MR}[1]{\mathrm{#1}}
\newcommand{\MC}[1]{\mathcal{#1}}

\newcommand{\bmx}{\bm{r}}
\newcommand{\bmv}{\bm{v}}

\newcommand{\kb}{k_\MR{B}}
\newcommand{\vth}{v_\MR{th}}
\newcommand{\partx}{\frac{\partial}{\partial \bmx}}
\newcommand{\partscx}[1]{\frac{\partial #1}{\partial x}}
\newcommand{\partv}{\frac{\partial}{\partial \bmv}}
\newcommand{\partt}{\frac{\partial}{\partial t}}

\newcommand{\Rbra}[1]{\left( #1 \right)}
\newcommand{\average}[1]{\left\langle #1 \right\rangle}
\newcommand{\partdiff}[2]{\frac{\partial #1}{\partial #2}}
\newcommand{\pdif}[2]{\frac{\partial #1}{\partial #2 } }

\begin{document}

\title{Anomalous transport phenomenon of a charged Brownian particle \\ under a thermal gradient and a magnetic field}

\author{Hiromichi Matsuyama}
 \email{matsuyama@r.phys.nagoya-u.ac.jp}
\author{Kunimasa Miyazaki}%
 \email{miyazaki@r.phys.nagoya-u.ac.jp}
\affiliation{Department of Physics, Nagoya University, Nagoya, 464-8602, Japan}

\date{\today}
\begin{abstract}
There is a growing interest in the stochastic processes
of nonequilibrium systems subject to non-conserved forces, such as the
magnetic forces acting on charged particles and
the chiral self-propelled force acting on active particles.
In this paper, we consider the stationary transport of non-interacting
 Brownian particles under a constant magnetic field in a
 position-dependent temperature background.
We demonstrate the existence of the Nernst-like stationary density current
perpendicular to both the temperature gradient and magnetic field,
induced by the intricate coupling between the non-conserved force and
the multiplicative noises due to the position-dependent temperature.
\end{abstract}

\maketitle


\section{Introduction}
\label{sec:level1}

In the past two decades, we have witnessed a tremendous development
of stochastic thermodynamic theory and transport phenomena of
systems far from equilibrium~\cite{seifert2012RepProg,sekimoto2010,klages2013nonequilibrium}.
Especially, the nonequilibrium systems in the presence of the non-conserved forces or the forces without time-reversal symmetry such have recently attracted a lot of attention~\cite{Chernyak2006jsmte,Chun2018,Tamaki2018,Souslov2019prl,Abdoli2020pre,Vuijk2020prr}.
The Lorentz force due to a magnetic field, the Coriolis force in the
rotating systems, the chiral self-propelled force of the active matters are
typical examples of the forces which violate the time-reversal symmetry.
The forces without time-reversal symmetry do not affect the stationary
distribution at the thermal equilibrium state, as exemplified by the
celebrated Bohr-van Leeuwen theorem for the classical systems in the magnetic field~\cite{Pradhan2010epl}.
Likewise, the linear response theory built upon the detailed balance condition is also unaltered by the presence of the time-anti-symmetric forces, as
Onsager's reciprocal relation claims~\cite{Onsager1931a,Onsager1931b}.
However, if
the system is in a nonequilibrium state beyond the linear response
regime,
 the violation of the detailed balance coupled with the violation of the time-reversal symmetry
can bring about anomalous transport phenomena and nontrivial stationary distribution~\cite{Souslov2019prl}.

In this paper, we demonstrate the simplest example of such anomalous transport,
 {\it i.e.}, a single classical Brownian motion in the presence of the temperature gradient and a constant magnetic field.
We show that the anomalous transport emerges perpendicular to both temperature gradient and magnetic field.

First, let us summarize what we know of a simple Brownian motion
under the temperature gradient and the magnetic field.
The force acting on a single Brownian particle moving with the velocity $\bmv$
is written as
\begin{equation}
\bF = -\zeta\bmv -\nabla \Phi(\bmx) - q\bB\times\bmv.
\label{eq:1-1}
\end{equation}
Here the first term is the drag force,
with the  friction coefficient $\zeta$,
the second is the force due to a conserved force
$\Phi(\bmx)$, and the last term is the Lorenz force,
where $q$ is the charge of the particle and $\bB$ is a magnetic field.
If the magnetic field is absent, the diffusion current is given by
$\bm{J}
= -\left(\zeta^{-1}\nabla \Phi + D\nabla\right)\rho
= -\zeta^{-1}\left(\nabla \Phi + \kb T\nabla\right)\rho$,
where $\rho=\rho(\bmx,t)$ is the probability distribution
function and $D=\kb T/\zeta$ is the diffusion coefficient.
If the magnetic field $\bB$ is turned on,
the velocity-dependent force $-\zeta\bmv$ is replaced by
$-(\zeta + q\bB\times)\bmv \equiv - \bzeta\cdot\bmv$ as seen in
Eq.~\eqref{eq:1-1}.
Here,
\begin{equation}
\bzeta = \zeta + q\bB\times
\label{eq:1-2zeta}
\end{equation}
is the friction matrix.
Therefore, the diffusion current is replaced by
\begin{equation}
\bm{J}
= -\bzeta^{-1}\cdot\left(\nabla \Phi + \kb T\nabla\right)\rho.
\label{eq:1-2}
\end{equation}
This relation holds even when $\bzeta$ depends on $\bmx$.
Eq.~\eqref{eq:1-2} looks deceptively simple but its derivation requires
careful adiabatic elimination of the velocity and
the importance of the asymmetric nature of the friction coefficient
should not be understated~\cite{Chun2018}.
Since the matrix coefficient, $\bzeta^{-1}$
in Eq.~\eqref{eq:1-2} contains
anti-symmetric components due to the Lorentz force,
the current perpendicular to both $\bB$ and  $\nabla \rho$ should be induced.
For example, if $\bB = (0, 0, B)$,  the friction matrix is written as
\begin{equation}
 \bzeta =\zeta \left(
\begin{array}{ccc}
 1 &  - \omega_s \tau& 0\\
 \omega_s \tau &  1 & 0\\
0              &  0 & 1\\
\end{array}
\right),
\label{eq:1-4a}
\end{equation}
where $\omega_s = qB/m$ is the cyclotron frequency and $\tau=m/\zeta$
is the relaxation time of the velocity.
$m$ is the mass of the Brownian particle.
Thus, the density gradient in the $x$-direction induces the
current in the $y$-direction that is proportional to
$J_y \propto \omega_s\tau \partial \rho/\partial x$.
This is nothing but the (classical) Hall effect.

Next, consider to place the system under the temperature gradient.
In this case, $T= T(\bmx)$ is spatially heterogeneous and
Eq.~\eqref{eq:1-2} should be modified as
\begin{equation}
\bm{J}
= -\bzeta^{-1}\cdot\left(\nabla \Phi + \nabla \kb T\right)\rho.
\label{eq:1-3}
\end{equation}
Note that $\kb T(\bmx)$ is now placed after $\nabla$.
This originates from the multiplicative nature of the position-dependent
random noises.
Eq.~\eqref{eq:1-3} was derived by an adiabatic elimination method in
Refs.~\cite{VanKampen1988,vanKampen1988jmp,Miyazaki2018pre}.
The presence of $\nabla T$ in Eq.~\eqref{eq:1-3}
implies that the temperature gradient can induce the density current
and it is the simplest example of the so-called Soret effect or
thermophoresis~\cite{degrootmazur1962}.
Eq.~\eqref{eq:1-3} also claims that the density current
induced by the temperature gradient can, in turn, induce
the Hall current.
The Hall effect due to the temperature gradient is referred to as the
Nernst effect~\cite{Pitaevskii2012book}.

What will happen if the system is confined by a wall or subject to the periodic
boundary condition in the direction of the temperature gradient, say,
along the $x$-axis when $\bB$ is exerted along the $z$-axis?
Can the Nernst effect induce the stationary density current
to the direction perpendicular to the $x$-axis?
Obviously, Eq.~\eqref{eq:1-3} claims that the answer is {\it No}, because the
density current in the $x$-direction, $J_x$, is prohibited by the
boundary condition in its direction.
If $J_x$ is absent, $J_y$ cannot be induced.
But this is not rigorously true.

In this paper, we analyze the Kramers equation of a single Brownian particle
in the presence of the temperature gradient and the magnetic field analytically
and demonstrate that the Nernst flow is induced when we go beyond the
linear response regime.
The Nernst effect is commonly observed in electronic or spin currents
or in rarefied plasma systems~\cite{Pitaevskii2012book,Behnia2016rpp}.
In the system we consider, this Nernst flow is explained in a subtle interplay between
the nonequilibrium condition (temperature modulation),
the non-conserved force (magnetic field), and the inertial effect of
the Brownian particles.
The magnitude of the current is found to be linear in $\bB$ and the
higher-order derivative of the temperature, $\nabla^3 T$.
In the terminology kinetic theory, it is the Burnett corrections~\cite{toda1991, hansen2013}.
We also verify the result by a Brownian Dynamics simulation
which quantitatively agrees with the analytical results up to the lowest
order in $\bB$.

This paper is organized as follows.
In Sec.~\ref{sec:sec2}, we start with the Kramers equation for
the velocity and position of a Brownian particle and carry out the
standard inverse-friction expansion~\cite{risken1996fokker}.
The solution is given in the form of the Brinkman's hierarchy up to the
lowest order in the coupling of the temperature gradient and the
magnetic field~\cite{Brinkman1956, Durang2015}.
In Sec.~\ref{sec:sec3}, we derive the solution of the hierarchy and
compare the results of the Brownian Dynamic simulations.
We provide semi-quantitative argument on our results in Sec.~\ref{sec:sec4}.
In Sec.~\ref{sec:sec5}, we conclude.

\section{Inverse-Friction expansion \label{sec:sec2}}

We start with the underdamped Langevin equation for a non-interacting
Brownian particle in the presence of the magnetic field $\bB$ and the position-dependent temperature
profile $T(\bmx)$.
The Langevin equation is given by
\begin{equation}
m\dot{\bmv} = -\zeta \bmv -\nabla \Phi(\bmx) - q{\bB}\times \bmv  + \sqrt{2\zeta\kb T(\bmx)}~\bmeta,
\label{eq: Underdamped_Langevin0}
\end{equation}
where $m$, $\zeta$, $q$ are the mass, friction coefficient, and the
charge of the particle, respectively.
$\kb$ is the Boltzmann constant and
$-\nabla \Phi(\bmx)$ is the conserved force.
$\bmeta(t)$
is the random white noise which satisfies
$\average{\eta_i(t)}=0$,
and $\average{ \eta_i(t)\eta_j(t)} =\delta_{ij}\delta(t-t'),$ with
$i, j = x,y,z$.
We carry out the standard inverse-friction expansion method of the Brownian system under the condition that all parameters in Eq.~\eqref{eq: Underdamped_Langevin0},
$T$, $\zeta$, and $B$ are functions of the position $\bmx$.
By expanding the Kramers equation to the lowest order in the inverse of the friction coefficient, one obtains the overdamped Langevin
equations~\cite{risken1996fokker, Kaneko1981}
in the presence of the thermal
gradient\cite{VanKampen1988,vanKampen1988jmp,Durang2015} and the magnetic
field \cite{Chun2018}.
We need to explore the higher-order in the inverse of the friction to
obtain the coupling of the temperature gradient and the magnetic field.
The higher-order expansion was carried out for one-dimensional system under the temperature
gradient~\cite{Widder1989physica, Stolovitzky1998pla}.
We generalize the method to the case that both the magnetic field and the
temperature gradient are present.

For simplicity, we assume that the magnetic field is exerted in the
$z$-direction, $\bB = (0,0, B)$, and $\bB$ is independent of $z$, so
that we consider the two-dimensional Brownian motion in the $(x,y)$-plane.
Eq.~\eqref{eq: Underdamped_Langevin0} 
is written as,
\begin{equation}
\dot{\bmv} = {\bm f} - {\bf G}\cdot \bmv  + \sqrt{2\gamma k_B T/m} ~\bmeta,
\label{eq: Underdamped_Langevin}
\end{equation}
where ${\bm f}= -m^{-1}\nabla \Phi$ and $\gamma=\zeta/m$.
${\bf G}= \bzeta/m$ is the friction matrix with $\bzeta$ defined by
Eq.~(\ref{eq:1-4a}).
The Kramers equation for the probability distribution function for
a single Brownian particle $P(\bmv,\bmx,t)$
corresponding to Eq.~\eqref{eq: Underdamped_Langevin} is
written as
\begin{equation}
\partt P(\bmv,\bmx,t)
= \MC{L}P(\bmv,\bmx,t)
= (\MC{L}_\MR{rev} + \MC{L}_\MR{irr})P(\bmv,\bmx,t), \\
\label{eq:Kramers}
\end{equation}
where
$\MC{L}_\MR{rev}$ and $\MC{L}_\MR{irr}$ are the reversible and
irreversible operators, respectively, defined by
\begin{equation}
\left\{
\begin{aligned}
&\MC{L}_\MR{rev} = -\bmv\cdot \partx -{\bm f}(\bmx) \cdot \partv - \frac{q}{m} \partv \cdot \Rbra{\bmv \times {\bm B}}, \\
&\MC{L}_\MR{irr} = \gamma \partv \cdot \Rbra{\bmv + \vth^2 \partv},
\end{aligned}
\right.
\end{equation}
where $\vth = \sqrt{\kb T(\bmx)/m}$ is the thermal velocity.
The position-dependence of $\vth$ through $T=T(\bmx)$ will play
an essential role in the following derivation.
It is convenient to introduce the new operator by
\begin{equation}
\begin{split}
\bar{\MC{L}}
&\equiv  \Rbra{\rho(v_1)\rho(v_2)}^{-1/2} \MC{L}  \Rbra{\rho(v_1)\rho(v_2)}^{1/2},
\end{split}
\label{eq:barL}
\end{equation}
with $\rho(v) =(2\pi \vth^2)^{-1} e^{-v^2/2\vth^2}$
and rewrite Eq.~\eqref{eq:Kramers} as
\begin{equation}
\partt \bar{P}(\bmv,\bmx,t) = (\bar{\MC{L}}_\MR{rev} + \bar{\MC{L}}_\MR{irr}) \bar{P}(\bmv,\bmx,t),
\label{eq:Kramers2}
\end{equation}
where $\bar{P}(\bmv,\bmx,t) \equiv  P(\bmv,\bmx,t)/\sqrt{\rho(v_x)\rho(v_y)}$.
Eq.~\eqref{eq:barL} transforms $\MC{L}_\MR{irr}$ to Hermitian operator given by
\begin{equation}
\bar{\MC{L}}_\MR{irr} = -\gamma \bm{b}^{\dagger} \cdot \bm{b},
\end{equation}
where we introduced the
ladder operator $\bm{b}$ and $\bm{b}^\dagger$ by
\begin{equation}
\begin{split}
\bm{b} = \vth \partv + \frac{\bmv}{2\vth}, \quad \bm{b}^{\dagger} =  -\vth \partv + \frac{\bmv}{2\vth} .
\end{split}
\end{equation}
The $n$-th orthogonal eigenfunctions  of
$-\gamma {b}_i^{\dagger} {b}_i$ ($i=x,y$) are given by
\begin{equation}
\begin{split}
\psi_n (v_i ,\bmx) &= \frac{ (b^{\dagger}_i)^n \psi_0(v_i) }{\sqrt{n!}} = \frac{\psi_0(v_i)}{\sqrt{n!2^n}}\ \MR{H}_n\Rbra{\frac{v_i}{\sqrt{2}\vth}}
\end{split}
\label{eq:def_psi}
\end{equation}
with $\psi_0(v) \equiv \sqrt{\rho(v)}$.
Here $n$ is the non-negative integer and
$\MR{H}_n(z)$ are the Hermite polynomials.
Note that the $\bmx$-dependence comes through $T(\bmx)$ in $\vth$.
The orthogonal eigenfunctions of $\bar{\MC{L}}_\MR{irr}$ are written as
$\Psi_{k, l}(\bmv, \bmx) = \psi_{k}(v_1,\bmx) \psi_{l}(v_2,\bmx)$.
One can construct the solution of Eq.~\eqref{eq:Kramers2} by expanding
$\bar{P}(\bmv,\bmx,t)$ in terms of the eigenfunctions
$\Psi_{k,l}(\bmv ,\bmx)$ as
$\bar{P}(\bmv,\bmx,t) =\sum_{k,l = 0}^{\infty} c_{k,l}(\bmx,t) \Psi_{k,l}(\bmv, \bmx),$
where $c_{k,l}(\bmx,t)$ is the expansion coefficients.
Rewriting $\bar{\MC{L}}_\MR{rev}\bar{P}(\bmv,\bmx,t)$ in terms of the ladder
operator is a straightforward task.
After tedious calculations, one arrives at the Kramers equation in
terms of the ladder operator given by
\begin{equation}
\begin{split}
&
\sum_{k,l = 0}^{\infty} \pdif{c_{k,l}}{t}\Psi_{k,l}
\\
&
= -\sum_{k,l = 0}^{\infty} \Biggl[\gamma c_{k,l} {\bm b}^{\dagger} \cdot{\bm b} + \Rbra{{\bm D}c_{k,l}}
\cdot {\bm b}  + \Rbra{\hat{\bm D} c_{k,l}} \cdot {\bm b}^{\dagger}  \\
&+ c_{k,l}  \partdiff{\vth}{\bmx} \cdot ({\bm b} + {\bm b}^{\dagger})
\left\{ {\bm b}^{\dagger} \cdot ({\bm b} + {\bm b}^{\dagger})\right\}-
 c_{k,l} {\bm \omega}_s \cdot ( {\bm b}^{\dagger} \times {\bm b})
 \Biggr]
\\
&
\times \Psi_{k,l},
\end{split}
\label{eq:Kramers3}
\end{equation}
where ${\bm \omega}_s= \omega_s {\bm B}/|{\bm B}|$ and
the operator $\bm{D}$ and $\hat{\bm{D}}$ are defined by
\begin{equation}
\begin{split}
\bm{D}&= \vth\partx, \quad \hat{\bm{D}} = \vth \partx  - \frac{\bm{f}}{\vth}.
\end{split}
\end{equation}
From Eq.~\eqref{eq:Kramers3}, one can construct the
hierarchical recurrence equation for $c_{k,l}(\bmx, t)$ using
the properties of eigenfunctions and the ladder operator
\begin{equation}
b_i \psi_{k} = \sqrt{k}\psi_{k -1},
~~
b_i^{\dagger}\psi_{k} = \sqrt{k +1}\psi_{k+1}.
\end{equation}
Using the orthogonal relation of the eigenfunctions of $\psi_n (v_i ,\bmx)$,
we obtain the hierarchy equation for $c_{k,l}$ given by
\\
 \begin{widetext}
 \begin{equation}
 \hspace{-1.5cm}
 \begin{split}
 &
  \pdif{c_{k,l}}{t}
 =
 -\gamma (k + l)c_{k,l} -\sqrt{k+1} D_xc_{k+1,l}
 - \sqrt{l+1}D_yc_{k,l+1} - \sqrt{k} \hat{D}_x c_{k-1,l}
 -\sqrt{l}\hat{D}_y c_{k,l-1} \\
 &
 + \sqrt{k(l+1)}~ \omega_s c_{k-1,l+1}- \sqrt{(k+1)l}~ \omega_s c_{k+1,l-1}
-\partdiff{\vth}{x}\left\{(k+1)^{\frac{3}{2}}c_{k+1,l} + 2k^{\frac{3}{2}}c_{k-1,l} + \sqrt{k(k-1)(k-2)}c_{k-3,l}\right\}
 \\
 &-\partdiff{\vth}{y}\left\{(l+1)^{\frac{3}{2}}c_{k,l+1} + 2l^{\frac{3}{2}}c_{k,l-1} + \sqrt{l(l-1)(l-2)}c_{k,l-3}\right\}
 \\
 &-\partdiff{\vth}{x}\left\{\sqrt{k+1}lc_{k+1,l} + \sqrt{(k+1)l(l-1)}c_{k+1,l-2} +\sqrt{k}lc_{k-1,l} + \sqrt{kl(l-1)}c_{k-1,l-2}\right\}
 \\
 &
 -\partdiff{\vth}{y}\left\{\sqrt{l+1}kc_{k,l+1}+ \sqrt{(l+1)k(k-1)}c_{k-2,l+1} +\sqrt{l}kc_{k,l-1} + \sqrt{lk(k-1)}c_{k-2,l-1}\right\}. \\
 \end{split}
 \label{eq:hierarchy_eq}
 \end{equation}
 \end{widetext}
This equation is a generalized version of the Brinkman's hierarchy~\cite{Brinkman1956}.
Up to here, the equation is rigorous and contains no approximation.
The hierarchy equation is already derived in Ref.~\cite{Chun2018} for
${\bm B}\neq 0$ (but with a constant $T$) and
in Ref.~\cite{Durang2015} for the $\bmx$-dependent $T(\bmx)$ (but in the
absence of ${\bm B}$).

\subsection{Lowest order term}

Let us first collect the lowest order term in $\MC{O}(\gamma^{-1})$ in Eq.~\eqref{eq:hierarchy_eq}.
The terms that survive are $c_{k,l}$ which satisfy $k+l = 0,1$ and
$\partial c_{k,l}/\partial t$ which satisfy $k+l =0$.
The set of equations for $c_{0,0}$ and ${\bm c}_1 \equiv
(c_{1,0},c_{0,1})$ is given by
\begin{equation}
\left\{
\begin{aligned}
&
\pdif{c_{0,0}}{t}
=-{\bm D}\cdot{\bm c}_1 - \partdiff{\vth}{\bmx}\cdot {\bm c}_1 =-\partx \cdot  (\vth{\bm c}_1),
\\
&
0 = - {\bf G}(\bmx)\cdot{\bm c}_1-\Rbra{\hat{\bm D}  + 2\partdiff{\vth}{\bmx}}c_{0,0},
\end{aligned}
\right.
\label{eq:hierarchy_eq0}
\end{equation}
where ${\bf G}(\bmx)$ is defined by Eq.~\eqref{eq: Underdamped_Langevin}.
The probability distribution for the position, $\rho(\bmx, t)$ is related to the
coefficient $c_{0,0}$ by
\begin{equation}
\rho(\bmx, t) = \int_{-\infty}^{\infty}d \bmv~ P(\bmv,\bmx,t) = c_{0,0}(\bmx, t).
\end{equation}
By substituting the second equation of Eq.~\eqref{eq:hierarchy_eq0} to
the first, we obtain the Smoluchowski equation
\begin{equation}
\begin{split}
 \pdif{\rho}{t}
&
=\partx \cdot \bzeta^{-1} \cdot\Rbra{\pdif{\Phi}{\bmx} + \partx\kb T}\rho,
\end{split}
\label{eq:Smoluchowski0}
\end{equation}
where $\bzeta^{-1} = m^{-1}{\bf G}^{-1}$ is the inverse friction matrix
which is defined by Eq.~\eqref{eq:1-2zeta}.
The probability flux
\begin{equation}
\begin{split}
{\bm J} \equiv  - \bzeta^{-1}\cdot\Rbra{\pdif{\Phi}{\bmx} + \partx\kb T}\rho
\label{eq:flux0}
\end{split}
\end{equation}
is identical to Eq.~\eqref{eq:1-3} which was originally derived
by adiabatic elimination of the momentum variable directory from the
Langevin equation~\cite{VanKampen1988,vanKampen1988jmp,Miyazaki2018pre}.
We address again that Eq.~\eqref{eq:Smoluchowski0} corresponds to the
overdamped Langevin equation and as
discussed in Introduction, it fails to predict the stationary current if
the system is confined in the direction of the temperature gradient.

\subsection{Beyond the lowest order term}

Next, we go beyond the linear order in $\gamma^{-1}$ and consider
the higher-order terms in the hierarchy equation of Eq.~\eqref{eq:hierarchy_eq}.
Since the calculations are involved, we shall consider only the
stationary state where  $\partial c_{k,l}/\partial t = 0$.
We also assume, for simplicity, that
the temperature is modulated only in the $x$-direction so that $T(\bmx) = T(x)$,
the magnetic field is constant and does not depend on $\bmx$ so that
${\bm B}(\bmx)={\bm B} = (0,0,B)$, $\zeta$ is also constant, and finally there is no external
conserved force so that $\Phi(\bmx)  = 0$.
We shall impose the wall or periodic boundary condition in the
direction of the temperature gradient at $x=0$ and $L$, so that
$\average{v_x(x)} = 0$ for arbitrary position $x$.
This is equivalent to impose $c_{1,0} = 0$ for arbitrary order in $\gamma^{-1}$.
We also assume that the system is not bounded in the $y$-direction.
The probability distribution does not depend on $y$ due to the symmetry and therefore
all coefficients $c_{k,l}$ are functions of $x$.
Then the hierarchy equation, Eq.~\eqref{eq:hierarchy_eq} is simplified
as
\begin{widetext}
\begin{equation}
\hspace{-1.5cm}
\begin{split}
0 =
&
-\gamma (k + l)c_{k,l}
- D_x\left(\sqrt{k+1} c_{k+1,l} - \sqrt{k} c_{k-1,l} \right)
+ \omega_s  \left\{\sqrt{k(l+1)}c_{k-1,l+1}-\sqrt{(k+1)l}c_{k+1,l-1}\right\}
\\
&
-\pdif{\vth}{x}\left\{(k+1)^{\frac{3}{2}}c_{k+1,l} + 2k^{\frac{3}{2}}c_{k-1,l} + \sqrt{k(k-1)(k-2)}c_{k-3,l}\right.
\\
&
\left.
+
\sqrt{k+1}lc_{k+1,l} + \sqrt{(k+1)l(l-1)}c_{k+1,l-2} +\sqrt{k}lc_{k-1,l} + \sqrt{kl(l-1)}c_{k-1,l-2}
\right\}.
\end{split}
\label{eq:hierarchy_sec3}
\end{equation}
\end{widetext}
The question is whether $\average{v_y(x)}$, or the probability flux $J_{y }(x)\equiv \rho(x)\langle v_y(x) \rangle$,
is finite in the nonequilibrium condition even when $\average{v_x(x)} = 0$.
$J_{y}(x)$ is written as
\begin{equation}
\begin{split}
J_{y }(x)
  &=\int v_y P(\bmv,x)d\bmv.
\end{split}
\end{equation}
Let us calculate this density current by applying the Orthogonality of the Hermite polynomial.
The current is written in terms of $c_{0,1}(x)$ as
\begin{equation}
\hspace{-1cm}
\begin{split}
J_{y}(x)
 = \int v_y c_{0,1}(x) \Psi_{0,1}(\bmv) \Psi_0(\bmv)d\bmv
&=\vth c_{0,1}(x),
\end{split}
\label{eq:def_avevy}
\end{equation}
where we used the definition for the eigenfunction, Eq.~\eqref{eq:def_psi}.
The hierarchy equation, Eq.~\eqref{eq:hierarchy_sec3}, gives the relation between $c_{0,1}$ and $c_{1,1}$;
\begin{equation}
\begin{split}
\gamma c_{0,1} =& - D_xc_{1,1}  -2\partscx{\vth}c_{1,1}. \\
\end{split}
\label{eq:c_01andc_11}
\end{equation}
We need to pick up the terms which survive at the same order of $\gamma^{-1}$
to calculate a non-zero $c_{1,1}$.
Let us denote $c_{k,l}^{(l)}$ for the coefficients of the order of $\gamma^{-l}$.
We already showed $c_{0,0}^{(0)}$ satisfies the Smoluchowski equation,
Eq.~\eqref{eq:Smoluchowski0}.
Next order terms of the order of $\gamma^{-1}$ are
\begin{equation}
\begin{split}
c_{3,0}^{(1)} = -\frac{\sqrt{6}}{3\gamma}{\partscx{\vth}} c_{0,0}^{(0)},~~~
c_{1,2}^{(1)} = -\frac{\sqrt{2}}{3\gamma}{\partscx{\vth}} c_{0,0}^{(0)},
\end{split}
\end{equation}
which are derived from Eq.~\eqref{eq:hierarchy_sec3}.
These terms represent the kinetic energy current parallel to thermal gradient.
The kinetic energy current calculated from these terms gives the same result as the linear irreversible thermodynamics.
The next order terms of $\MC{O}(\gamma^{-2})$ are
$c_{2,0}^{(2)},c_{0,2}^{(2)},c_{2,1}^{(2)}$, which are written, using
$c_{0,0}^{(0)}$,  as
\begin{equation}
\left\{
\begin{aligned}
c_{2,0}^{(2)} =& -\frac{\sqrt{3}}{2\gamma} \bar{D}_3c_{3,0}^{(1)}
= \frac{\sqrt{2}}{2\gamma^2} \bar{D}_3 \partscx{\vth} c^{(0)}_{0,0},
\\
c_{0,2}^{(2)} =& -\frac{1}{2\gamma} \bar{D}_3 c_{1,2}^{(1)} =
 \frac{\sqrt{2}}{6\gamma^2} \bar{D}_3 \partscx{\vth}c^{(0)}_{0,0},
\\
c_{2,1}^{(2)} =& \frac{1}{3\gamma}\Rbra{ 2\omega_s c_{1,2}^{(1)} -\sqrt{3}\omega_s c_{3,0}^{(1)}} = \frac{\sqrt{2}\omega_s}{9\gamma^2}  \partscx{\vth}c^{(0)}_{0,0},
\end{aligned}
\right.
\label{eq:c_2nd}
\end{equation}
where we introduced the differential operator
\begin{equation}
\bar{D}_3 = D_x + 3\partscx{\vth}.
\end{equation}
$c_{1,1}^{(3)}$ is the order of $\MC{O}(\gamma^{-3})$ and given by
\begin{equation}
\begin{split}
\sqrt{2}\gamma c_{1,1}^{(3)} =& -\bar{D}_3c_{2,1}^{(2)} +\omega_sc_{0,2}^{(2)}-\omega_sc_{2,0}^{(2)}.
\label{eq:c11_3rd}
\end{split}
\end{equation}
Substituting Eq.~\eqref{eq:c_2nd} to Eq.~\eqref{eq:c11_3rd},
$c_{1,1}^{(3)}$ is written in terms of $c^{(0)}_{0,0}$ as
\begin{equation}
\begin{split}
c_{1,1}^{(3)} =
-\frac{4\omega_s}{9\gamma^3}  \bar{D}_3\partscx{\vth} c^{(0)}_{0,0}.
\end{split}
\end{equation}
The correction of $c_{0,1}$ of $\MC{O}(\gamma^{-4})$ is written
using the relation between $c_{0,1}$ and $c_{1,1}$,
Eq.~\eqref{eq:c_01andc_11}, as
\begin{equation}
\begin{split}
c_{0,1}^{(4)}
=& \frac{2\omega_s}{9\gamma^4} \frac{N^{(0)}}{\vth} \frac{\partial^3}{\partial x^3} \vth^2,
\label{eq:c01}
\end{split}
\end{equation}
where $c^{(0)}_{0,0} = N^{(0)}/\vth^2$ and $N^{(0)} = (\int dx/\vth^2(x))^{-1}$ is the normalization constant for $c_{0,0}^{(0)}(x)$ to satisfy $\int c^{(0)}_{0,0} dx= 1$.
Eq.~(\ref{eq:c01}) is the first non-zero contribution of $c_{0,1}$.
From Eq.~(\ref{eq:def_avevy}), we arrive at
\begin{equation}
\begin{split}
J_{y}(x) =\frac{2\omega_sN^{(0)}}{9\gamma^4} \partdiff{^3}{x^3} \frac{\kb T(x)}{m}.
\end{split}
\label{eq:ydensity_flux}
\end{equation}
This is the main result of this paper.
It shows that the Nernst-like current appears as the super Burnett
term~\cite{toda1991, hansen2013} and
is the lowest order term which is finite and it is linear in $|\bm{ B}|$.

\section{Comparison with the numerical simulation}
\label{sec:sec3}

We verify the analytical expression, Eq.~\eqref{eq:ydensity_flux},
by the direct numerical simulation.
We simulate the Langevin equation given by
Eq.~\eqref{eq: Underdamped_Langevin0} by the standard Brownian Dynamics simulation for a
two-dimensional system.
We apply a constant magnetic field along the $z$-direction, $\bm{B} = (0,0,B)$.
Therefore, the motion of the non-interacting Brownian particles along the $z$-direction is
decoupled with the motion on the $(x,y)$-plane.
As we have detailed in the previous section, the anomalous current is
proportional to the third derivative of the temperature profile and
therefore we consider a system where the periodic temperature modulation imposed
along the $x$-direction, which is given by
\begin{equation}
\begin{split}
T(x) = T_\MR{ave}\left\{1 + \alpha\cos\Rbra{\frac{2\pi x}{L}}\right\},
\end{split}
\label{eq:temperature_simu}
\end{equation}
where $\alpha$ is the amplitude of the temperature modulation that controls
the strength of the
temperature gradient and $T_{\MR{ave}}$ is the mean temperature.
The Brownian motion of particles is affected by the local temperature through
the random force $\sqrt{2\zeta \kb T(x)}~\bmeta$ in Eq.~\eqref{eq: Underdamped_Langevin0}.
In our simulation, we set the system size $L$ and the velocity relaxation time
$\tau\equiv \gamma^{-1}$ as the unit of length and time.
We introduce the non-dimensional parameter for the magnetic field and
temperature by
\begin{equation}
B^* = \omega_\MR{s}\tau, ~~~~T^* = \frac{\kb T_\MR{ave}}{m}\frac{\tau^2}{L^2},
\end{equation}
where $\omega_\MR{s}=qB/m$ is the cyclotron frequency.
We adopt the periodic boundary condition for a two dimensional box of
the size $L$.
We perform the Brownian Dynamics simulation for $N = 4,194,304$ non-interacting particles in the
simulation box and take $1,000$ samples for each simulation run.
Since we consider the non-interacting particles, the probability
density $\rho(x,t)$ can be interpreted as the density field.
To obtain the current profile of $J_{y}(x)$ and $J_{x}(x)$  as a function of
$x$, we divide the system into 25 bins in the $x$-direction and then
average over $y$-direction.
We choose simulation time step of $dt = 10^{-4} \times \min(2\pi/ B^*,1/\sqrt{T^*})$
which is much shorter than relevant time scales.
Sampling time is $\tau_\MR{sample} = \max(2\pi/ B^*,1/\sqrt{T^*})$.

\begin{figure}[htb]
\centering
  \includegraphics[width=0.9\columnwidth,angle=-0]{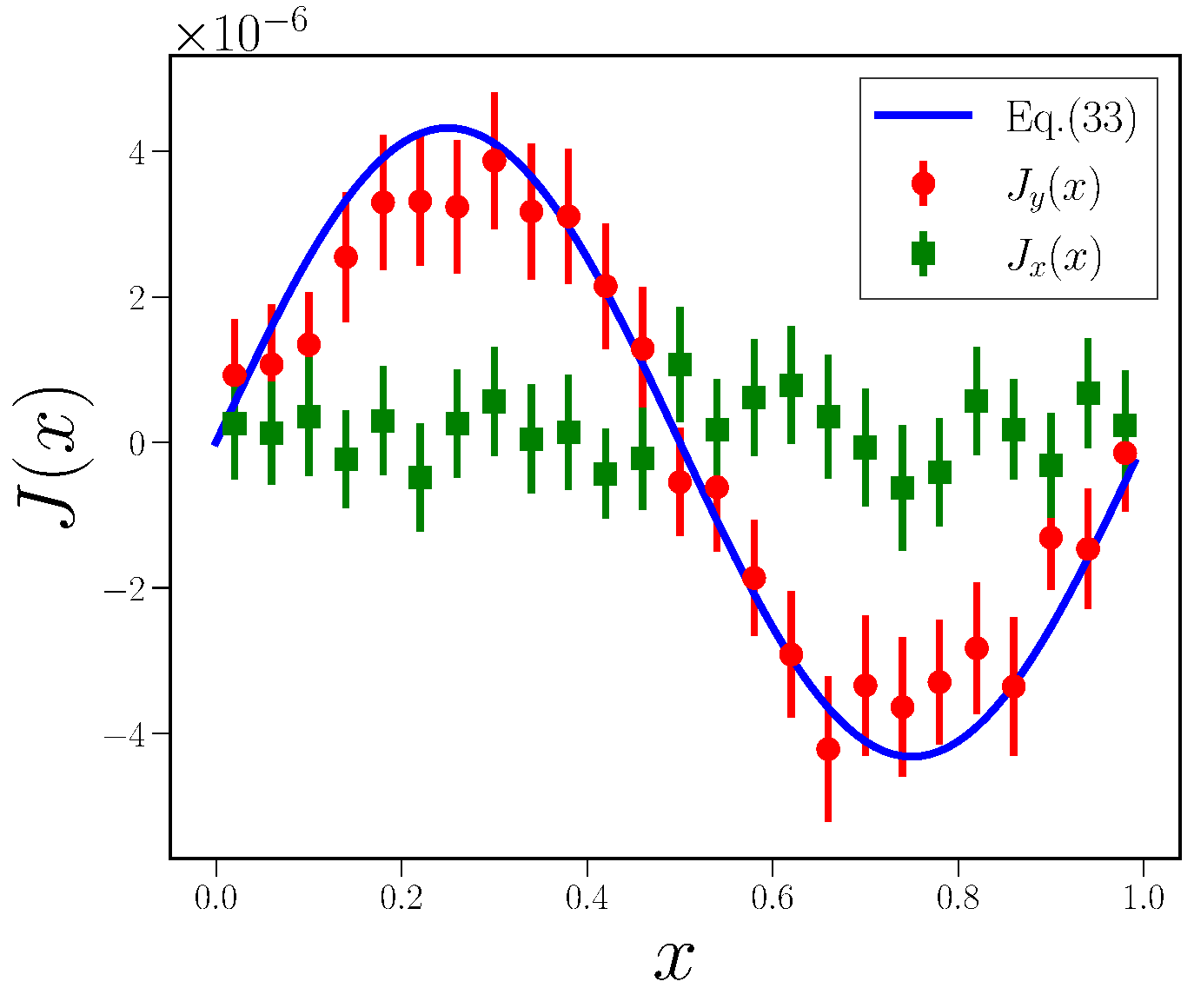}
  \caption{The average density current to $x$ and $y$-direction.
Theoretical result Eq.~\eqref{eq:ydensity_flux} (Blue solid line)
and the simulation result of the current for $x$-direction $J_{x}(x)$ (Red dots) and
$y$-direction $J_{y}(x)$ are shown.
The error bar for the simulation result is the standard error.
The parameters are $\alpha=0.2,T^*=0.002,B^*=0.1$.
 }
\label{fig:comp_theory_simu}
\end{figure}

Figure~\ref{fig:comp_theory_simu} shows the $x$-dependence of the
current
$J_{y}(x)$ along the $y$-direction,
predicted by our theory, Eq.~\eqref{eq:ydensity_flux}, and obtained by the simulation .
We chose the parameters $\alpha=0.2$, $T^*=0.002$, and $B^*=0.1$.
The simulation result of the density current for $y$-direction agrees very well with the theoretical prediction.
Note that the density current in the $x$-direction $J_{ x}(x)$
vanishes as expected.
Since we have truncated the expansion up to the linear order in
$\omega_{\MR{s}}$, we chose a relatively small value of $B^*$.

\begin{figure}[htb]
\centering
  \includegraphics[width=0.9\columnwidth,angle=-0]{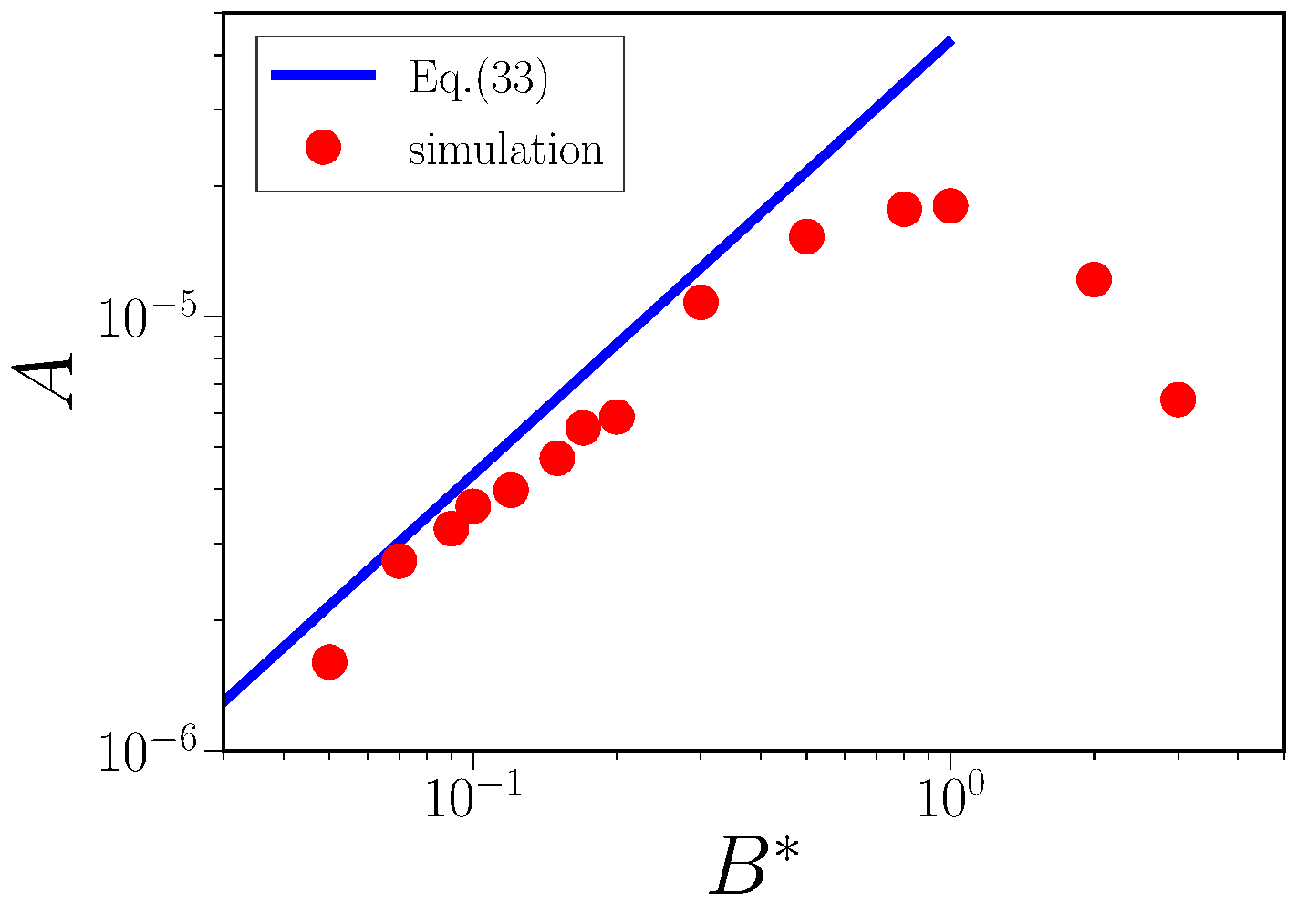}
  \caption{The $B$-dependence of the current intensity for the theoretical
 result (Blue solid line) and the simulation result (Red dots).
Simulation parameters are $\alpha=0.2,T^*=0.002$.
}
    \label{fig:B_dependence_simu}
\end{figure}

Next, we investigate the $B$-dependence of the current.
The current has the sinusoidal profile fitted by $A\sin(2\pi x)$ and
we plot the $B$ dependence of the amplitude $A$
in Figure \ref{fig:B_dependence_simu} together with the theoretical
prediction.
The theoretical prediction is $A = {16\pi^3\alpha N^{(0)}T^*B^*}/{9\gamma^4}$ from Eq.~\eqref{eq:ydensity_flux}.
In the small $B$-regime, the simulation result agrees well with the
theoretical prediction.
As $B$ increases, the simulation result reaches a peak and then
decreases.
Note that the magnetic field for which the cyclotron frequency
becomes comparable with the inverse of the velocity relaxation time,
$\omega_s \tau = 1$, is at $B^{\ast} \approx 1$.
The value agrees with the position of the peak of $A$ shown in Fig.~\ref{fig:B_dependence_simu}.
This implies that when the magnetic field is strong, the
cyclotron frequency is very fast and the cyclotron radius becomes
small, which reduces the bend of the particle trajectories, as is the
case for the Nernst coefficient for the rarefied
gases~\cite{Pitaevskii2012book}.
This nonlinear effect is not taken into account in our theoretical analysis
(Eq.~\eqref{eq:ydensity_flux})
because we retain only the linear terms in $B$.

{\section{semi-quantitative argument}
\label{sec:sec4}}

We provide a semi-quantitative argument about how the anomalous transport
arises and why the current depends on $\partial_x^3 T$.
Let us consider the hydrodynamic description of the Brownian particles.
In the stationary state, the frictional force due to the current and the
osmotic pressure should be balanced.
If the magnetic field is present, the Lorentz force should be also
added, so that we have
\begin{equation}
 0 =  - \nabla p -\zeta \bm{J} - q \bB \times \bm{J},
\end{equation}
where $p$ is the osmotic pressure field and $\bm{J}$ is the velocity
field (and not the velocity of respective particles).
Inverting the equation, we obtain the solution for $J_y$ given
by
\begin{equation}
\begin{aligned}
J_{y}(x)  \approx - \frac{\omega_s \tau}{\zeta} \pdif{p}{x}
\end{aligned}
\label{eq:vy2}
\end{equation}
up to the linear order in $B$.
For the non-interacting Brownian particles, the osmotic pressure is
given by $p = \rho \kb T$.
This is exact if the local equilibrium condition is satisfied.
We assume that $p$ is still proportional to $\rho T$ even when the local equilibrium
condition is violated.
We expand $\rho$ as $\rho(x) = \rho^{(0)}(x) + \rho^{(1)}(x) + \cdots $ in the
$1/\gamma$-expansion.
As we have shown  in Section~\ref{sec:sec2} (see Eq.~\eqref{eq:flux0}),
$\rho^{(0)}(x)T(x)=$constant for the lowest order term in the steady state in the case that the conserved force is absent.
This solution corresponds to the overdamped approximation~\cite{Miyazaki2018pre}.
The next order terms in $1/\gamma$ is $\rho^{(2)}$ ($= c_{0,0}^{(2)}$ in
Section~\ref{sec:sec2}) .
$\rho^{(2)}$ satisfies
\begin{equation}
\begin{split}
\Rbra{\partscx{} + \frac{2}{\vth}\partscx{\vth}}\rho^{(2)}
=
-\sqrt{2}\Rbra{\partscx{} + \frac{2}{\vth}\partscx{\vth}}c^{(2)}_{2,0}.
\end{split}
\end{equation}
We have already derived $c^{(2)}_{2,0}$ in Section~\ref{sec:sec2} (see Eq.~\eqref{eq:c_2nd}).
The solution of this equation is
\begin{equation}
\rho^{(2)} = - \frac{N^{(0)}}{2T(x)\gamma^2} \frac{\partial^2 T(x)}{\partial x^2}+\frac{m N^{(0)}}{\kb T(x)}N^{(2)}.
\label{eq:density_2nd}
\end{equation}
where $N^{(0)}$ is a normalization constant for $c^{(0)}_{0,0}$ introduced at Eq.~\eqref{eq:c01} and $N^{(2)} = \sqrt{2}\int  c^{(2)}_{2,0}  dx$ is a normalization constant
which is determined in such a way that
$\int \rho^{(2)} dx = 0$.
It is noteworthy that Eq.~\eqref{eq:density_2nd} does not depend on $B$
and is present in the absence of $B$.
Combining the above results, we conclude that the osmotic pressure
gradient is not zero but instead it is given by
\begin{equation}
\begin{split}
\pdif{p}{x}
&= \pdif{}{x} \rho^{(2)}(x) \kb T(x)
= - \frac{\kb N^{(0)}}{2\gamma^2} \frac{\partial^3 T(x)}{\partial x^3}.
\end{split}
\label{eq:effective_pressure}
\end{equation}
Combining this expression with Eq.~\eqref{eq:vy2}, we arrive at
\begin{equation}
\begin{split}
J_{y}(x) =\frac{\omega_s N^{(0)}}{2\gamma^4} \frac{\partial^3 }{\partial x^3} \frac{\kb T(x)}{m} \\
\end{split}
\end{equation}
Aside from a numerical factor, this is identical to our final
result, Eq.~\eqref{eq:ydensity_flux}.
We conclude that the small deviation of the density field from the local
equilibrium induces the local pressure modulation,
$\partial p/\partial x\neq 0$ which causes the local velocity field in
the $x$-direction and eventually leads to the velocity field in the
$y$-direction.
The current in the $x$-direction is only local and
net current integrated over $x$-direction should vanish.
\\

\section{Discussion and Conclusions}
\label{sec:sec5}

We have shown that there exists a stationary density current of
non-interacting Brownian particles
perpendicular to the temperature modulation and the magnetic
field, even when the system is confined in the direction of the
temperature modulation by the wall or the periodic boundary condition.
This Nernst-like effect is proportional to the magnetic field and the
third derivative of the temperature modulation.
This has been explained by analytically solving the Kramers equation by
the standard inverse friction expansion.
These effects are beyond the linear response regime in that the current is
due to the super Burnett correction or the higher-order derivative of the temperature gradient.
This anomalous current can not be explained in the overdamped limit, which
corresponds to the lowest order contribution in the $1/\gamma$
expansion, at which the local equilibrium assumption is valid.

It is not easy practically to observe the anomalous current in
realistic systems because the large magnetic field and the small
friction coefficient are required.
However, it is important to recognize that the coupling between the force
which does not satisfy the time-reversal symmetry can induce the
macroscopic current, if small, in the simple classical system.
In active matter systems,
the non-conserved force which violates the time-reversal symmetry combined
with the strong nonequilibrium conditions is ubiquitous~\cite{VanTeeffelen2008,Lowen2016,Souslov2019prl,Abdoli2020pre,Vuijk2020prr,Yang2021prl, huang2021pnas}.
The coupling of the effect of the inertia and the nonequilibrium
conditions is also known to bring about the nontrivial macroscopic
effect in active matter systems, in stark contrast with the equilibrium
Brownian motion where the short time scales are relevant in the underdamped equation
does not affect the macroscopic behaviors~\cite{Mandal2019prl, Lowen2020jcp}.
The result in this paper demonstrates the counter-intuitive effects
due to a subtle interplay between
the non-conserved force (the Lorentz force),
the nonequilibrium condition (the inhomogeneous temperature),
and the inertia effect (the finite $m$) in possibly the simplest model system.
Considering the coupling of these effects in realizable systems such as active matter systems is a promising future direction.

\begin{acknowledgments}
The authors would like to thank Shuji Tamaki and Keiji Saito for inspiring this work and fruitful discussion.
This research is supported by the Japan Society for the Promotion of Science (JSPS) KAKENHI
(No.~19H01812   
and  20H00128).   
\end{acknowledgments}

\end{document}